\documentclass[aps, pra, twocolumn, notitlepage,superscriptaddress
]{revtex4-2}
\usepackage{amssymb,latexsym,amsmath}
\usepackage{bm}
\usepackage[normalem]{ulem}
\usepackage{amssymb,amsmath}
\usepackage{graphicx,xcolor}
\usepackage[titletoc, title]{appendix}
\usepackage[caption=false,lofdepth,lotdepth]{subfig}
\usepackage{hyperref}   
\usepackage{soul}
\hypersetup{%
	pdfpagemode=FullScreen,
	pdfstartpage=1,
	pdfmenubar=true,
	pdftoolbar=true,
	colorlinks = true,
	linkcolor=blue,
	citecolor=blue,
	urlcolor=blue,
	bookmarksopen=false
}

\begin{document}

\title{Making ghost vortices visible in two-component Bose-Einstein condensates}
\author {Andrii Chaika}
\affiliation{Department of Physics, Taras Shevchenko National University of Kyiv, 64/13, Volodymyrska Street, Kyiv 01601, Ukraine}
\author{Andrea Richaud}
\email[Corresponding author: ]{andrea.richaud@upc.edu}
\affiliation{Scuola Internazionale Superiore di Studi Avanzati (SISSA), Via Bonomea 265, I-34136, Trieste, Italy}
\affiliation{Departament de F\'isica, Universitat Polit\`ecnica de Catalunya, Campus Nord B4-B5, E-08034 Barcelona, Spain}
\author {Alexander Yakimenko}
\affiliation {Department of Physics, Taras Shevchenko National University of Kyiv, 64/13, Volodymyrska Street, Kyiv 01601, Ukraine}

\affiliation { Dipartimento di Fisica e Astronomia ’Galileo Galilei’, Università di Padova, via Marzolo 8, 35131 Padova, Italy }
 \affiliation {
            Istituto Nazionale di Fisica Nucleare, Sezione di Padova, via Marzolo 8, 35131 Padova, Italy }

\date{\today}

\begin{abstract}
Ghost vortices constitute an elusive class of topological excitations in quantum fluids since the relevant phase singularities fall within regions where the superfluid density is almost zero. Here we present a platform that allows for the controlled generation and observation of such vortices. Upon rotating an imbalanced mixture of two-component Bose-Einstein condensates (BECs), one can obtain necklaces of real vortices in the majority component whose cores get filled by particles from the minority one. The wavefunction describing the state of the latter is shown to harbour a number of ghost vortices which are crucial to support the overall dynamics of the mixture. Their arrangement typically mirrors that of their real counterpart, hence resulting in a ``dual" ghost-vortex necklace, whose properties are thoroughly investigated in the present paper. We also present a viable experimental protocol for the direct observation of ghost vortices in a ${}^{23}\mathrm{Na}$ $+$ ${}^{39}\mathrm{K}$ ultracold mixture. Quenching the inter-component scattering length, some atoms are expelled from the vortex cores and, while diffusing, swirl around unpopulated phase singularities, thus turning them directly observable.

\end{abstract}

\maketitle

\section{Introduction}
Vortices are the most fascinating and universal structures in classical and quantum fluids. 
The appearance of vortices in a quantum fluid constitutes the hallmark of superfluidity.  Unlike a classical fluid, in fact, a superfluid cannot rotate as a rigid body due to the irrotational character of its velocity field, which is, in turn, proportional to the gradient of the phase field associated to the macroscopic wavefunction \cite{Leggett2001}. For this reason, a superfluid can gain angular momentum only in the form of discrete topological defects (vortices).    

Quantum vortices are found in various physical systems, ranging from liquid Helium \cite{Donnelly,Volovik1988} to quantum gases \cite{Matthews1999,Madison2000,Anderson2000,AboShaer2001}, but including also superconductors \cite{Blatter1994}, exciton-polariton condensates \cite{Lagoudakis2008}, and quantum fluids of light \cite{Carusotto2013}. Typically, a quantum vortex corresponds to a singularity in the phase field (i.e. a point around which the phase ``rolls up" from 0 to $2\pi$) and, for the energy to be bounded, to a density depression. More specifically, in a mean-field description, the velocity profile around a vortex diverges as $r^{-1}$ (where $r$ is the distance from the phase singularity) and the density profile goes as $r^2$ \cite{Pitaevskii2016}. Hence, vortices are often pictured as funnel-like holes in otherwise uniform density plateaus, around which the quantum fluid exhibits a swirling flow. 

The typical diameter of vortex cores greatly depends on the properties of the superfluid and ranges from $\sim 10^{-10}$ m for superfluid Helium-4 to $\lesssim 10^{-6}$ m for atomic BECs \cite{Pethick2008,Keepfer2020}, length scales that are (considerably) smaller than the optical resolution of imaging apparatuses. To circumvent this intrinsic limitation and achieve a direct visualization of vortex lines, clever stratagems have been designed, such as the injection of small particles of solid hydrogen into samples of liquid Helium (some of them indeed get trapped along vortex filaments) \cite{Bewley2006} (see also Ref. \cite{Giurato2020}), or the use of two-component BECs where one component fills the core of the vortex in the other component (and thus enlarges it because of the inter-component repulsion) \cite{Anderson2000}. As regards single-component quasi two-dimensional (2D) BECs, given the submicron size of vortex cores, imaging protocols often rely on a period of ballistic expansion of the atomic cloud prior to image acquisition \cite{Anderson2010,Neely2010}, even though recent \emph{in situ} imaging protocols \cite{Wilson2015,Kwon2021} have opened the door to the direct observation of vortices in contexts where BEC expansion is either impractical or would alter the vortex features. 

Soon after the first experimental observations of vortices in atomic BECs \cite{Matthews1999,Madison2000,AboShaer2001}, the dynamics of vortex lattice formation upon trap rotation was extensively investigated by means of numerical simulations of 2D Gross-Pitaevskii (GP) equations \cite{Tsubota2002,Kasamatsu2003}. In this context, a new class of topological excitations termed ``ghost vortices" was pointed out. The latter are phase singularities that appear at the outskirt of the atomic cloud when it is rotated. They, by gradually approaching the Thomas-Fermi boundary (from outside), give place to surface waves which ultimately result in dynamical instability. When this happens, i.e. for sufficiently high rotation frequencies, the phase singularities enter the atomic cloud, form a lattice, and start to give a non-vanishing contribution to the total angular momentum of the system. 

The concept of ``ghost vortex", i.e. that of a phase defect in regions where the superfluid is almost absent, also recurs in other physical systems. In Ref. \cite{Wen2010}, for example, ``hidden vortices" were found along the central barrier of a rotating double-well potential and, since they carry angular momentum, were shown to be the crucial ingredient for the Feynman rule \cite{Feynman1955} to be satisfied.  In Refs. \cite{Fujimoto2010,Sabari2018}, ghost vortices were found in the oscillating potential which is used to nucleate vortex dipoles in its wake, while in Ref. \cite{Griffin2020JJ} they resulted from the snake instability of solitons and preluded vortex turbulence. Interestingly, in a many-body description of trapped BECs, it is possible to recognize the presence of ``phantom vortices", topological defects of the spatial coherence, but not of the density \cite{Weiner2017}. Ghost vortices play a crucial role also in some recent technological applications, where coherent matter waves can be conveniently employed to realize high-precision accelerometers \cite{Liu2019,Bland2022}, and to characterize the current-phase relationship of quantum devices \cite{Eckel2014,Abad2015,Gallemi2015}.  

The presence of ghost vortices is not limited to single-component BECs. Indeed, they can be found in Bose-Bose \cite{Wen2013} and in Bose-Fermi \cite{Wen2014} mixtures, where they are generated by collective excitations and the Landau instability \cite{Wu2001} with negative excitation frequency. They are also found in dipolar BECs \cite{Sabari2018}, in mixtures featuring the Rashba-Dresselhaus spin-orbit coupling \cite{Yang2019,Su2020}, in the presence of an inhomogeneous artificial gauge field \cite{Hejazi2020}, and in self-bound quantum droplets \cite{Examilioti2020}.

In all these systems, ghost vortices support the nucleation of the usual (i.e. real) vortices in the bulk-density region. Be them in the periphery of a rotating condensate \cite{Tsubota2002,Kasamatsu2003,Wen2013,Wen2014} or in the interior of a moving obstacle \cite{Fujimoto2010,Sabari2018}, ghost vortices provide the seeds of topological defects which can eventually enter the high-density regions.  

In this work, we focus on a class of systems where ghost vortices not only are present but play a crucial role in the overall dynamics: two-component BECs in the immiscible regime, where one component hosts quantum vortices and the other component fills vortex cores. The dynamical properties of these ``massive vortices"  \cite{Richaud2020,Griffin2020Magnus,Richaud2021,Richaud2022rk,Richaud2022Collisions} are determined by the interplay of superfluid Vortex Dynamics \cite{Sonin1987,Barenghi2001,Kim2004,Fetter2009} and Newtonian (inertial) physics, and disclose intriguing phenomena such as the presence of cyclotron-like orbits \cite{Richaud2021}, precession reversal \cite{Richaud2022rk}, and two-vortex collisions \cite{Richaud2022Collisions}. 

While the aforementioned works were concerned with the \emph{dynamics} of massive vortices, here we shift the focus onto the properties of the minority wavefunction, the one associated with the core-filling component, both the manifest and the hidden ones. In particular, we show that the uniform motion of precession exhibited by vortex necklaces \cite{Hess1967,Campbell1979,Yarmchuk1979,Kim2004,Barry2015,Cawte2021} naturally comes with the formation of ghost vortices in the minority component. These phase singularities in regions where the relevant superfluid density is (almost) zero directly originates from the irrotational character of the velocity field and support the precession of the massive cores, integral with the hosting vortices.

Our paper is organized as follows.
Section \ref{sec:Rotatig_two_component_BEC} introduces the physical system, summarizes the effective point-vortex model, and provides analytical results about the precession frequency of vortex necklaces in the presence of core mass, as well as a detailed energetic stability of such configurations. Section \ref{sec:Ngos_of_ghost_vortices} then studies the arrangement of ghost vortices in the minority component and highlights their crucial role in the overall dynamics of the Bose-Bose mixture. To this purpose, analytical results from the massive point-vortex model are validated against numerical simulations of coupled GP equations. Section \ref{sec:GPE_Quenches} presents a viable experimental protocol to directly probe the existence of these elusive topological excitations. In the same spirit of standard fluid-dynamics experimental platforms, where colored smoke is injected to probe flow patterns \cite{Sieverding1983}, we release part of the minority component from the hosting vortices, and we observe it flowing around the original ghost vortices, thus turning them real. An indicator to quantify the degree of ``ghostliness" of a given phase defect is also presented, which can be easily employed to postprocess both numerical and experimental results. Section \ref{sec:Conclusions} is devoted to concluding remarks and outlook.

\section{Rotating two-component BEC}
\label{sec:Rotatig_two_component_BEC}
The dynamics of quasi-2D two-component BECs at zero temperature is described, at the mean-field level, by the two coupled GP equations
\begin{equation}
\label{eq:GPE_a}
i\hbar\frac{\partial \psi_a}{\partial t} = \left[-\frac{\hbar^2}{2m_a}\nabla^2 +V_{\rm tr} + g_{aa} |\psi_a|^2 + g_{ab}|\psi_b|^2 \right]\psi_a
\end{equation}
\begin{equation}
\label{eq:GPE_b}
i\hbar\frac{\partial \psi_b}{\partial t} = \left[-\frac{\hbar^2}{2m_b}\nabla^2 +V_{\rm tr} + g_{ab} |\psi_a|^2 + g_{bb}|\psi_b|^2 \right]\psi_b,
\end{equation}
where $\psi_a=\sqrt{\rho_a}e^{i\theta_a} $ and $\psi_b=\sqrt{\rho_b}e^{i\theta_b}$ are the order parameters associated to the two condensed components, $m_a$ and $m_b$ are the two atomic masses, and $g_{ij}=\sqrt{2\pi}\hbar^2 a_{ij}/(m_{ij} d_z)$ are the effective intra- and inter-species interactions in 2D. Parameters $m_{ij}=1/{(m_i^{-1}+m_j^{-1})}$ constitute the effective masses, $a_{ij}$ the intra- and inter-species $s$-wave scattering lengths, and $d_z$ represents the effective thickness of the quasi-2D atomic cloud. The macroscopic wavefunctions $\psi_a$ and $\psi_b$ are normalized to $N_a$ and $N_b$, the number of particles in the two condensed components, respectively. The trapping potential is such that $V_{\rm tr}(r)=0$ if $r<R$ and $V_{\rm tr}(r)=V$ if $r>R$, where $V$ should be sufficiently larger than $\mu_a$, so to ensure confinement of the system within a disk of radius $R$ ($\mu_a$ is the chemical potential associated to component-$a$). In our simulations, we employed $V\sim 10^4\,\mu_a$. 

\begin{figure}[h!]
    \centering    \includegraphics[width=1\columnwidth]{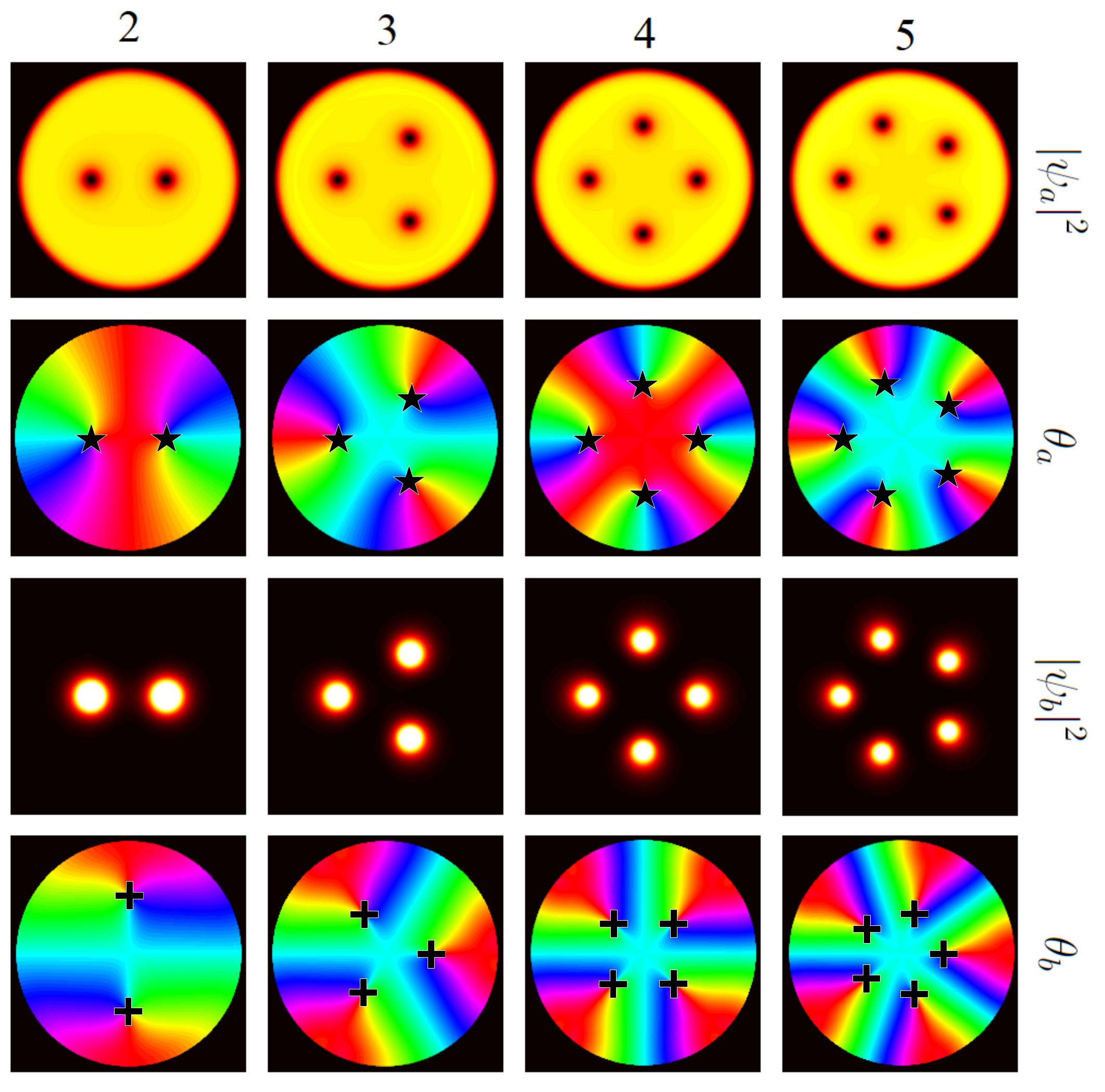}
    \caption{Regular $N$-gons of massive vortices in a hard-wall circular trap (each column corresponds to $N=2,\,3,\,4,\,5$). As customary, vortices in $\psi_a$ correspond to density holes in $\rho_a$ (first row) and singularities (black stars) in the associated phase field $\theta_a$ (second row). As expected, the density distribution of the filling component, $\rho_b$, features peaks (third row) at the centers of component-$a$ vortices. Less expectedly, singularities (black crosses) are present in the phase field $\theta_b$ (fourth row) in regions where $\rho_b$ is vanishingly small ($\sim 10^{3}$ times smaller than the peak density): these are {\it ghost vortices}. The properties of ghost-vortex arrays are described in Sec. \ref{sec:Ngos_of_ghost_vortices}, while a convenient experimental protocol to turn ghost vortices into real vortices is presented in Sec. \ref{sec:GPE_Quenches}. The following parameters have been used: $N_a=5 \times 10^4$, $N_b=1.5 \times 10^3$,  $R=50\, \mu m$, $m_a=3.82 \times 10^{-26}$ kg,  $m_b=6.48 \times 10^{-26}$ kg, $g_a=52 \times (4 \pi \hbar^2 a_0)/{m_a}$, $g_b=7.6 \times (4 \pi \hbar^2 a_0)/{m_b}$, $g_{ab}=24.2 \times (2 \pi \hbar^2 a_0)/{m_{ab}}$, $l_z=2\, \mu m$. Each $N$-gon has different angular velocity: $\Omega=5.5$ rad/s for $N=2$, $\Omega=6.5$ rad/s for $N=3$, $\Omega=7.5$ rad/s for $N=4$, $\Omega=8.5$ rad/s for $N=5$.}
    \label{fig:Ghost_vortices_arrangement}
\end{figure}

Equations (\ref{eq:GPE_a}) and (\ref{eq:GPE_b}) admit a class of solutions such that $\psi_a$ hosts one or more vortices and $\psi_b$ is localized at the cores of these vortices thus providing them with an effective inertial mass \cite{Richaud2020,Richaud2021}. The resulting ``massive vortices" represent composite topological excitations of the mixture \cite{Law2010,Richaud2020,Ruban2022}, and are stabilized by the assumed immiscibility $g_{ab}>\sqrt{g_{aa}\,g_{bb}}$ of the two components (if this condition is violated, a non-negligible mass can be still present in the vortex core, but it is not tightly localized therein \cite{Gallemi2018,Choudhury2022}). While considerable attention has been paid to their dynamics \cite{Richaud2020,Richaud2021,Richaud2022rk,Richaud2022Collisions}, important properties of the filling component $\psi_b$ are still unaddressed. In this article, we show that $\psi_b$ typically features ``ghost vortices", i.e. phase singularities in regions where the component-$b$ density, $\rho_b$, is vanishingly small. Moreover, we point out their fundamental role in the overall dynamics of the system, and propose a viable experimental protocol to turn them into \emph{real} vortices, so to gain a more direct insight into their features and properties. To this purpose, we consider systems of few massive vortices confined in a hard-wall circular trap, a type of trapping potential which is available in modern experimental apparatuses \cite{Navon2021} and which allows one to investigate physical phenomena without worrying about those undesired density-gradient effects inherently present with traditional harmonic traps. More specifically, we focus on regular configurations where $N$ massive vortices constitute the vertices of a regular $N$-gon  (see Fig. \ref{fig:Ghost_vortices_arrangement}). This choice allows us to precisely determine the structure of the associated ghost-vortex array and thus allows for a direct comparison with experiments. 

\subsection{Rotating $N$-gons of massive vortices} 
\label{sub:Rotating_Ngons}
In the absence of the filling component (i.e. for $N_b=0$) it is well known that various $N$-gons of \emph{real} (hence the subscript ``$r$") vortices, differing in the value of $N$, in the radius $r_r(N)$ of the circle where the \emph{real} vortex $N$-gon is inscribed in, as well as in the presence or absence of an extra vortex at the center of the circular hard-wall trap, get energetically (meta)stable depending on the value of the angular frequency $\Omega$ at which the system is rotated (see also Sec. \ref{sub:Energetic_Stability}). We recall that the aforementioned vortex $N$-gons are stationary in the rotating reference frame, while, in the laboratory reference frame, they rotate at the same frequency as the external rotation $\Omega$ \cite{Kim2004}. 

The presence of core mass alters the standard functional dependence $\Omega_N(r_r)$, representing the angular frequency at which a vortex $N$-gon of radius $r_r$ rotates. The reason is that, in the presence of a non-zero core mass, each vortex can be subject to an additional Magnus-like force \cite{Sonin1997,Griffin2020Magnus}. 


A particularly convenient analytical tool to estimate the dynamics of massive quantum vortices, as well as the properties of rotating $N$-gons of massive vortices is represented by the massive point vortex model \cite{Richaud2020,Richaud2021,Richaud2022Collisions,Richaud2022rk}, which generalizes previous point-vortex models \cite{Kim2004} where the possible presence of core mass was neglected. In the framework of a time-dependent variational approximation (which neglects compressibility-related effects, such as sound waves propagation), in fact, superfluid Vortex Dynamics and Newtonian inertial Physics can be treated on equal footing and the resulting effective point-like Lagrangian
\begin{equation}
\label{eq:Lagrangian_pointlike}
 L= \sum_{j=1}^N \left[ \frac{M_j}{2}\dot{\bm{r}}_j^2 + \frac{k_j\rho_a}{2} (\dot{\bm{r}}_j \times \bm{r}_j \cdot \hat{z}) \frac{r_j^2-R^2}{r_j^2}\right] - V,
\end{equation}
where
$$
     V=\frac{\rho_a}{4\pi}\sum_{j=1}^N k_j^2 \ln \left(1-\frac{r_j^2}{R^2}\right)+
$$
\begin{equation}
\label{eq:V}
     \frac{\rho_a}{4\pi}\sum_{i<j} k_ik_j \ln  \left(\frac{R^2 - 2\bm r_i\cdot\bm r_j  + r_i^2r_j^2/R^2}{r_i^2 -  2\bm r_i\cdot\bm r_j  + r_j^2}\right),
\end{equation}
was shown to well capture the trajectories $\{\bm{r}_j(t)\}$ of interacting massive vortices \cite{Richaud2020,Richaud2021,Richaud2022rk,Richaud2022Collisions}, bypassing the need of solving the actual GP equations (\ref{eq:GPE_a}) and (\ref{eq:GPE_b}). The meaning of model parameters in Lagrangian (\ref{eq:Lagrangian_pointlike}) is the following: $M_j=N_b m_b/N$ is the mass of each core (we assume that component-$b$ bosons are equally subdivided among the $N$ vortices), $k_j=\pm h/m_a$ is the strength of the $j$-th vortex, while $\rho_a=m_a n_a = m_a N_a /(\pi R^2)$ is the planar mass density of component-$a$ atoms, $R$ the radius of the trap. Lagrangian (\ref{eq:Lagrangian_pointlike}) includes a Newtonian inertial term $\propto M_j$ and a potential term which accounts for both inter-vortex interactions and the presence of a hard-wall circular potential which results, in turn, in the emergence of effective image vortices \cite{Kim2004}. 

\begin{figure}[h!]
    \centering    \includegraphics[width=1\columnwidth]{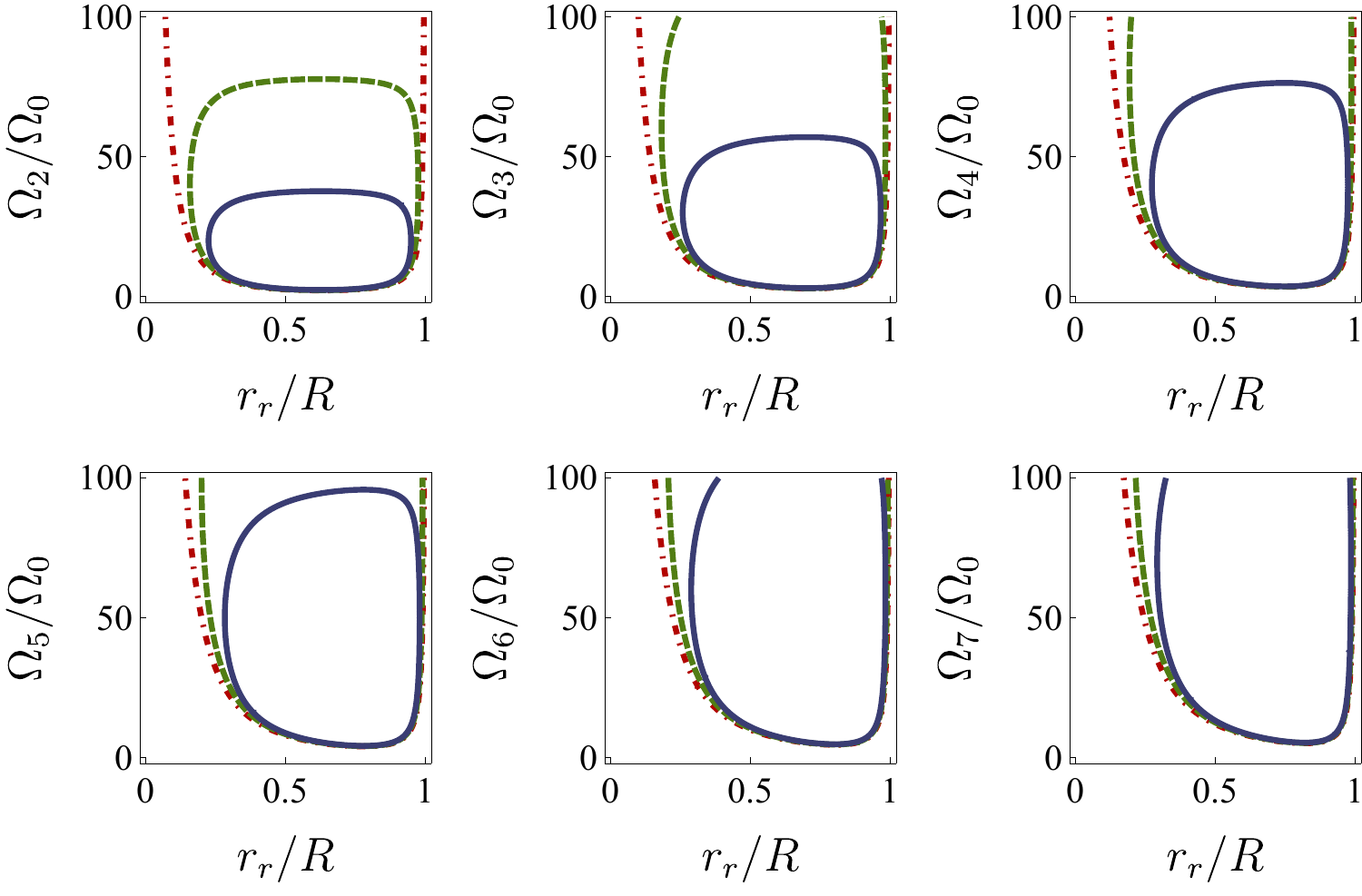}
    \caption{Plot of relation (\ref{eq:Precession_frequency_vs_r_r}). Each panel corresponds to a different value of $N$ (ranging from 2 to 7). Purple dot-dashed, green dashed, blue solid lines correspond, respectively, to $\mu=0$, $\mu=0.05$, and $\mu=0.10$.}
    \label{fig:Omega_N_vs_r_r}
\end{figure}

As already mentioned, in the presence of $N$ vortices, the motion equations associated to Lagrangian (\ref{eq:Lagrangian_pointlike}) feature a notable class of solutions where the $N$ vortices constitute the vertices of a regular $N$-gon of radius $r_r$, which rotates around its center at a rate $\Omega_N(r_r)$. Complex but straightforward calculations (see Appendix \ref{app:Potential_energy}) result into equation
\begin{equation} 
\label{eq:Precession_frequency_vs_r_r}
    \Omega_N\left(1-\frac{\mu}{2N} \Omega_N \right)= \frac{1}{r_r^2} \left( \frac{N-1}{2} + N \frac{r_r^{2N}}{1-r_r^{2N}}\right),
\end{equation}
where the precession frequency $\Omega_N$ is expressed in units of $\Omega_0=\hbar/(m_aR^2)$, the $N$-gon radius $r_r$ in units of the trap radius $R$, and $\mu=M_b/M_a=N_bm_b/(N_a m_a)=N M_j/(N_a m_a)$ is the dimensionless ratio between the total masses of the two components. The solutions of this equation are illustrated, for different values of $N$ and $\mu$, in Fig. \ref{fig:Omega_N_vs_r_r}.
One can observe that, for fixed $N$, upon increasing the value of $\mu$, the maximum possible precession frequency $\Omega_N$ decreases, while the allowed range of values for $r_r$ shrinks. Conversely, for fixed $\mu$, upon increasing the value of $N$, the maximum possible precession frequency $\Omega_N$ increases.

It is worth mentioning that Eq. (\ref{eq:Precession_frequency_vs_r_r}) admits solutions provided that 
\begin{equation} 
\label{eq:mu_inequality}
    \mu >\frac{N r_r^2 (1-r_r^{2N})}{N-1+(N+1)r_r^{2N}}.
\end{equation}
Upon determining the maximum of the right-hand side, for any given value of $N$, one can find a critical value  
\begin{equation}
    \label{eq:tilde_mu_N}
   \tilde{\mu}_N= \frac{N \left[N \left(\frac{N-1}{\sqrt{N^2-1}}-1\right)+1\right]^{\frac{1}{N}}}{\sqrt{N^2-1}}
\end{equation}
above which no solutions exist. For larger values of $\mu_N$, in fact, the vortices forming the $N$-gon are too massive to sustain any uniform circular precession, whatever the value of $r_r$. Quantity $\tilde{\mu}_N$ is minimum for $N=2$, at which it takes the value $\approx 0.45$, is a monotonically increasing function of $N$, and tends to $1$ for $N\to+\infty$. 

We conclude by remarking that the two-component BEC platform which we are going to present in Sec. \ref{sec:GPE_Quenches} features dimensionless mass ratios $\mu_N$ much smaller than the associated critical ratios (\ref{eq:tilde_mu_N}), so to ensure the existence of the corresponding massive-vortex $N$-gons.

\subsection{Energetic stability}
\label{sub:Energetic_Stability}
As anticipated in Sec. \ref{sub:Rotating_Ngons}, $N$-gons of massive vortices can be shown to constitute stable or metastable states of a two-component BEC trapped in a rotating potential. This property is particular important in view of an actual experimental realization because, in the presence of undesirable dissipative processes, the $N$-vortex system evolves so to minimize its free energy 
\begin{equation}
    \label{eq:Hamiltonian_rot}
    H^\prime_N(\{\bm{r}_j^\prime,\,\bm{p}_j^\prime\})= H_N(\{\bm{r}_j^\prime,\,\bm{p}_j^\prime\}) - \Omega \sum_{j=1}^N \bm{l}_j^\prime\cdot \hat{z}.
\end{equation}
The latter can be regarded as the energy of the $N$-vortex system in a frame rotating at frequency $\Omega$ \cite{Kim2004,Fetter2009} (in this frame, corresponding to the primed coordinates $\{\bm{r}_j^\prime,\,\bm{p}_j^\prime\}$, the vortex $N$-gon is at rest), while
\begin{equation}
    \label{eq:Hamiltonian}
    H_N= \sum_{j=1}^N \left[\frac{1}{2}M_j \dot{\bm{r}}_j^2 +\frac{\rho_a k^2}{4\pi}\ln\left(\frac{R}{\xi_a}\right) \right]+V
\end{equation}
is the $N$-vortex system's energy in the laboratory (unprimed) frame [$V$ is given by Eq. (\ref{eq:V}), while $\xi_a$ is the vortex-core-radius cutoff, corresponding to the characteristic width of the vortices] and 
\begin{equation}
    \label{eq:l_j}
    \bm{l}_j = \bm{r}_j \times \bm{p}_j = \frac{k\rho_a}{2}(R^2-r_j^2)\hat{z} + M_j \bm{r}_j \times \dot{\bm{r}}_j
\end{equation}
represents the canonical angular momentum associated to the $j$-th massive vortex, the linear momentum $\bm{p}_j$ being 
\begin{equation}
    \label{eq:p_j}
    \bm{p}_j=\frac{\partial L}{\partial \dot{\bm{r}}_j} = \frac{k_j\rho_a}{2} \frac{r_j^2-R^2}{r_j^2} \bm{r}_j \times \hat{z} + M_j \dot{\bm{r}}_j.
\end{equation}
Notice that $\bm{l}_j$ includes two contributions: one, $\propto \rho_a$, intrinsically originating from presence of the vortex, i.e. from the fact that each condensed boson in a vortex carries a non-zero angular momentum, the other, $\propto M_j$, ensuing from the presence of a moving core mass. We also observe, in passing, that energy (\ref{eq:Hamiltonian}) can be written in terms of the canonical variables
\begin{equation}
    \bm{z}=(\bm{r}_1,\,\bm{r}_2,\,\dots,\,\bm{r}_N,\,\bm{p}_1,\,\bm{p}_2,\,\dots,\,\bm{p}_N)^T
\end{equation}
in the laboratory frame leading to the equivalent form
\begin{equation}
    \label{eq:Hamiltonian_canonical}
    H_N(\{\bm{r}_j,\,\bm{p}_j\})= \sum_{j=1}^N \left[\frac{\left( \bm{p}_j -k_j \bm{A}_j \right)^2}{2M_j}  +\frac{\rho_a k^2}{4\pi}\ln\left(\frac{R}{\xi_a}\right) \right]+V
\end{equation}
where $\bm{A}_j=\frac{\rho_a}{2} \frac{r_j^2-R^2}{r_j^2}\bm{r}_j\times\hat{z}$ can be regarded as an effective vector potential \cite{Richaud2021,Richaud2022rk} [indeed this constitutes the reformulation, in the Hamiltonian framework, of the Lagrangian model (\ref{eq:Lagrangian_pointlike})]. Similarly, the Hamiltonian (\ref{eq:Hamiltonian_rot}) is written in terms of the canonical variables \begin{equation}
\label{eq:z_prime}
    \bm{z}^\prime=(\bm{r}_1^\prime,\,\bm{r}_2^\prime,\,\dots,\,\bm{r}_N^\prime,\,\bm{p}_1^\prime,\,\bm{p}_2^\prime,\,\dots,\,\bm{p}_N^\prime)^T
\end{equation}  
in the rotating reference frame.

Due to the symmetry characterizing regular $N$-gons of massive vortices, it is possible to write their energy (\ref{eq:Hamiltonian}) and angular momentum (\ref{eq:l_j}) in closed form (see Appendix \ref{app:Potential_energy}):
\begin{equation} \label{eq: Hamiltonian N-gon}
    \frac{H_N}{N}=\frac{\mu}{2N}\Omega^2 r_r^2 -\ln\xi_a + \ln\left(\frac{1-r_r^{2N}}{N\,r_r^{N-1}} \right)
\end{equation}
\begin{equation}
    \bm{l}_j\cdot\hat{z}= 1-r_r^2 + \frac{\mu}{N}\Omega r_r^2
\end{equation}
where the lengths $r_r$ and $\xi_a$ are expressed in units of $R$, the frequency $\Omega$ in units of $\Omega_0$, the angular momentum in units of $N_a\hbar$, and the energy in units of $N_a\hbar\Omega_0$.

\begin{figure}[h!]
    \centering    \includegraphics[width=1\columnwidth]{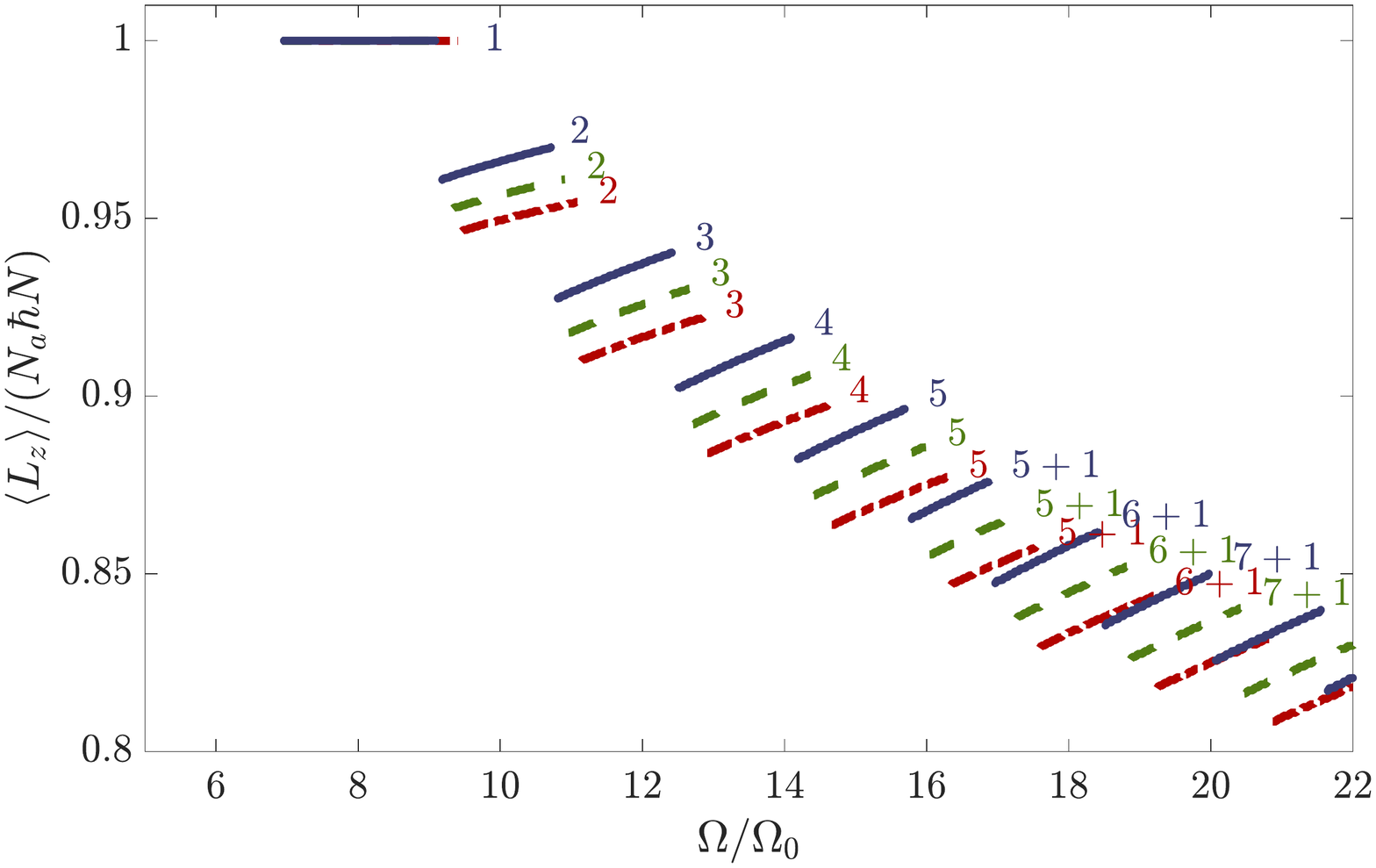}
    \caption{Plot of the total angular momentum of the minimum-energy configuration of the system as a function of the rotation frequency $\Omega$. Purple dot-dashed, green dashed, and blue solid lines correspond, respectively, to $\mu=0$, $\mu=0.05$, and $\mu=0.10$. Labels represent the value of $N$ associated to each curve segment. The extra label ``$+1$" corresponds to states where, besides the $N$ vortices constituting the vertices of the $N$-gon, there is an extra vortex at the trap center. Dimensionless ratio $\xi_a/R=10^{-3}$ has been assumed.}
    \label{fig:L_z_vs_Omega}
\end{figure}

Following the scheme of Ref. \cite{Kim2004}, we determined, for a rather extended range of rotation frequencies $\Omega$, the specific value of the $N$-gon radius $r_r^*$ which makes $H_N^\prime$ stationary by solving 
\begin{equation}
    \label{eq:Stationarity_condition}
    \frac{\partial H_N'(r_r)}{\partial r_r} =0.
\end{equation}
Then, among all these possible metastable states, we picked the one corresponding to the lowest free energy, $\min_{N}\{H^\prime_N\}$ and explicitly verified that it constitutes a local minimum of Hamiltonian (\ref{eq:Hamiltonian_rot}) in the phase space spanned by canonical variables (\ref{eq:z_prime}) (to be precise, a pair of eigenvalues of the Hessian Matrix should be neglected, as they are always zero because of the rotational symmetry of the Hamiltonian, which is also associated to angular-momentum conservation).
The result is illustrated in Fig. \ref{fig:L_z_vs_Omega}, in terms of the total angular momentum of the system,
\begin{equation}
    \label{eq:L_z}
    \langle L_z \rangle =\sum_{j=1}^N \bm{l}_j \cdot \hat{z},
\end{equation} 
for three different values of the dimensionless mass ratio $\mu$. We remark that quantity (\ref{eq:L_z}) is not only the sum of the $N$ canonical angular momenta (\ref{eq:l_j}), but it has also a more direct physical meaning, as it coincides with $\langle\hat{L}_{z,a}\rangle_{\psi_a}+\langle\hat{L}_{z,b}\rangle_{\psi_b}$, i.e. with the sum of the expectation values of the ($z$-component of the) angular momentum operator computed with respect to the time-dependent variational wavefunctions $\psi_a$ and $\psi_b$ used to derive the massive point vortex model (\ref{eq:Lagrangian_pointlike}) [see Eqs. (8)-(11) of Ref. \cite{Kim2004} and Eq. (15) of Ref. \cite{Richaud2021}] and which well approximate the actual condensates' wavefunctions.

As visible from Fig. \ref{fig:L_z_vs_Omega}, for each value of $\mu$, there is a set of critical frequencies $\{\Omega_{0,1}(\mu),\,\Omega_{1,2}(\mu),\,\dots\}$ at which the system's ground state transitions from an $N$-vortex configuration to a $(N+1)$-vortex configuration. For example, for rotation frequencies $\Omega<\Omega_{0,1}$, the lowest-free-energy state does not include any vortex, while, for $\Omega_{0,1}<\Omega<\Omega_{1,2}$, the $1$-vortex configuration is the globally stable one. As pointed out in Ref. \cite{Kim2004} for the massless case $\mu=0$, and as observed in an early experiment on rotating superfluid ${}^4\mathrm{He}$ \cite{Yarmchuk1979}, the system transitions from a configuration characterized by $N=5$ vortices in a ring (see label ``5" in Fig. \ref{fig:L_z_vs_Omega}) to a configuration featuring $N=5$ vortices in a ring plus one at the center (see label ``5+1" in Fig. \ref{fig:L_z_vs_Omega}). This is due to the fact that the $5+1$-vortex configuration has lower free energy (see Appendix \ref{app:Potential_energy}) than the $6$-vortex state (the latter corresponding to a vortex hexagon). Interestingly, our results confirm this property also in the presence of massive cores ($\mu\neq 0$), although the detailed set of critical frequencies is altered with respect to the massless case. More specifically, we observe that the presence of core mass \emph{anticipates} the $N$-to-$(N+1)$ transition, i.e. that $\Omega_{N,N+1}(\mu_>)<\Omega_{N,N+1}(\mu_<)$ for $\mu_> > \mu_<$. 

\section{$N$-gons of ghost vortices} 
\label{sec:Ngos_of_ghost_vortices}
\begin{figure}[h!]
    \centering    \includegraphics[width=1\columnwidth]{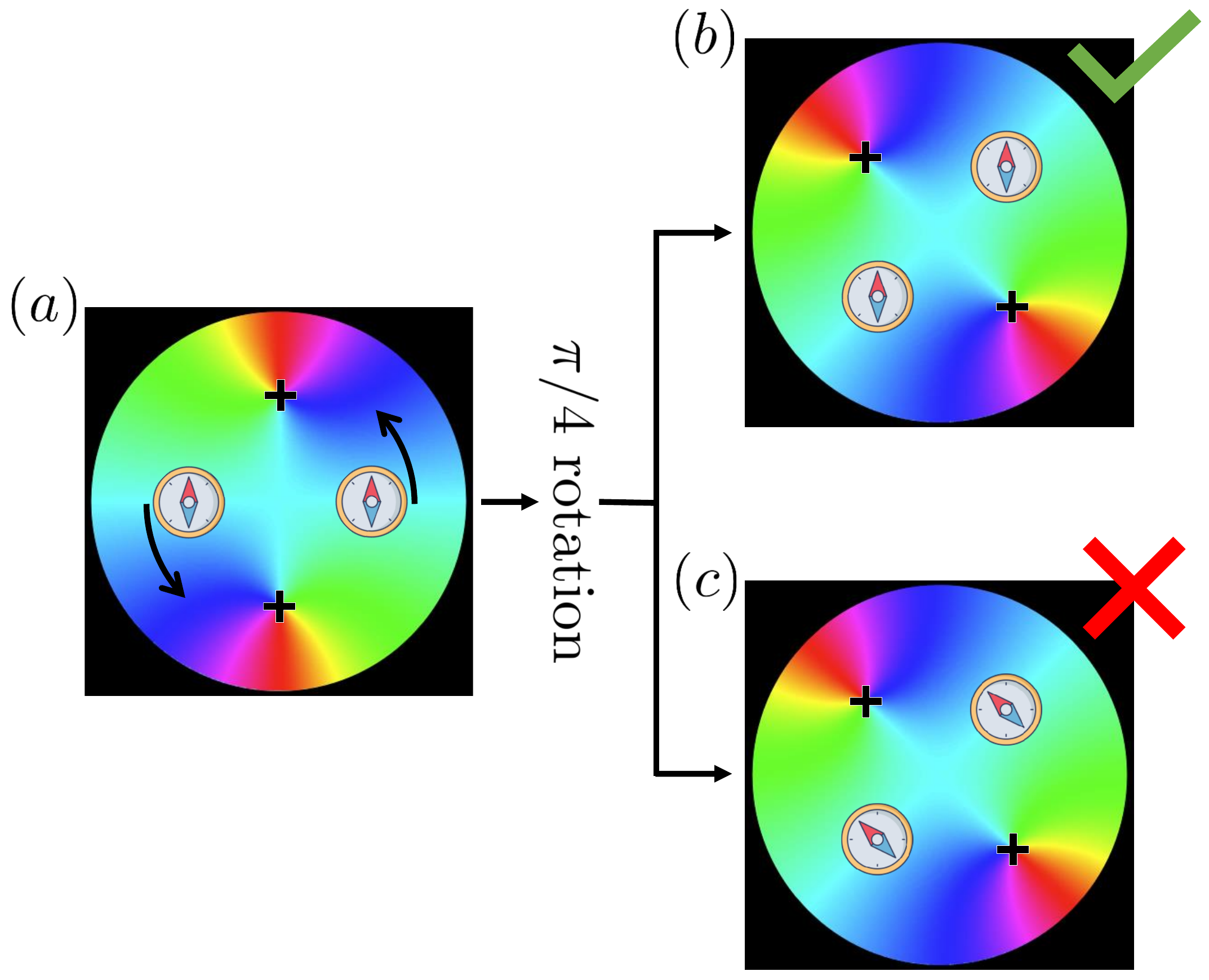}
    \caption{Schematic illustration of $N=2$ component-$b$ massive cores due to the presence of ghost vortices (black crosses) in $\psi_b$. The two compasses correspond to the two regions where the density $\rho_b$ is significantly different from zero. The orientation of the compass needles illustrates the irrotational character of the flow (\ref{eq:J_b}) associated with $\psi_b$. Starting, e.g., from an initial configuration [panel (a)] where the two massive cores lie on the horizontal diameter of the trap and the needles point upward, after an anticlockwise $\pi/4$-precession, the orientation of the massive cores is unaltered [panel (b)]. This is significantly different from a hypothetical rigid-body-like precession of the whole system [panel (c)].}
    \label{fig:Rotating_peaks}
\end{figure}

Having pointed out, in the previous section, the properties of massive-vortex necklaces and their robustness with respect to undesirable dissipation processes which may be present in an experimental platform, we now turn to the analysis of ghost vortices in the filling component. These structures originate from the properties of $\psi_b$, which is continuous and single-valued, and can be found in regions where the density $|\psi_b|^2$ is vanishingly small. As pictorially illustrated in panel (a) of Fig. \ref{fig:Rotating_peaks}, the presence of ghost vortices in $\psi_b=\sqrt{\rho_b}e^{i\theta_b}$ determines a non-zero phase gradient $\nabla\theta_b$, and hence a non-zero current density
\begin{equation}
    \label{eq:J_b}
    \bm{J}_b=m_b\rho_b \nabla\theta_b
\end{equation}
in regions where the density $\rho_b$ is non-zero. This is what supports the uniform precession of component-$b$ cores. We also notice that, since the flow associated with $\psi_b$ is irrotational, component-$b$ cores process around the trap's center but they do not rotate around their own axes [see panels (b) and (c) of Fig. \ref{fig:Rotating_peaks} for an intuitive representation and Ref. \cite{Richaud2020} for a quantitative analysis].

The number and the position of the ghost vortices in $\psi_b$ can be unambiguously determined by requiring that the tangential velocity $\bm{v}_N=\bm{\Omega}_N \times \bm{r}_r$ characterizing the uniform precession of a massive vortex in an $N$-gon (see Sec. \ref{sub:Rotating_Ngons}) matches the velocity vector field $\bm{v}_b=\hbar/m_b \nabla\theta_b$ evaluated at position $\bm{r}_r$. The resulting condition
\begin{equation}
    \label{eq:Precession_condition}
    \Omega_N r_r =\left.\frac{\hbar}{m_b} \nabla \theta_b\right|_{\bm{r}_r}\cdot \hat{\theta}
\end{equation}
(where $\hat{\theta}$ is the polar unit vector), together with simple symmetry-based considerations, allow one to determine the positions $\bm{r}_{g,1},\,\bm{r}_{g,2},\,\dots$ of the \emph{ghost} vortices (hence the subscript ``$g$", as opposed to the subscript ``$r$" used in Secs. \ref{sub:Rotating_Ngons} and \ref{sub:Energetic_Stability} to denote the positions of \emph{real} vortices) in $\psi_b$. 

To be more specific, we assume that the phase field $\theta_b$ is the superposition of $N_g$ ghost vortices, hence
\begin{equation}
\label{eq:theta_b}
    \theta_b(x,\,y) = \sum_{j=1}^{N_g}\arctan\left(\frac{y-\bm{r}_{g,j}\cdot \hat{y}}{x-\bm{r}_{g,j}\cdot\hat{x}}\right).
\end{equation}
In the light of the GP simulations which we performed (see Fig. \ref{fig:Ghost_vortices_arrangement}), one can argue that ghost vortices typically constitute the vertices of an $N_g$-gon (with $N_g=N$), which is rotated by an angle $\pi/N$ with respect to the $N$-gon of massive vortices, a sort of ``dual" necklace. Condition (\ref{eq:Precession_condition}) thus results into equation
\begin{equation}
    \label{eq:r_r_vs_r_g}
    \Omega = \frac{m_a}{m_b} \left[\frac{N}{{r_r}^2} \frac{\left(\frac{r_r}{r_g}\right)^{N} }{1+\left(\frac{r_r}{r_g}\right)^{N}}\right]
\end{equation}
where, again, frequency $\Omega$ is expressed in units of $\Omega_0$ and the lengths $r_r$ and $r_g$ in units of $R$. So, given an $N$-gon of massive vortices rotating at frequency $\Omega$ [see Eq. (\ref{eq:Precession_frequency_vs_r_r})], one can solve Eq. (\ref{eq:r_r_vs_r_g}) in the unknown $r_g$ and hence determine the position of ghost vortices in $\psi_b$.  
\begin{figure}
    \centering
    \includegraphics[width=1\columnwidth]{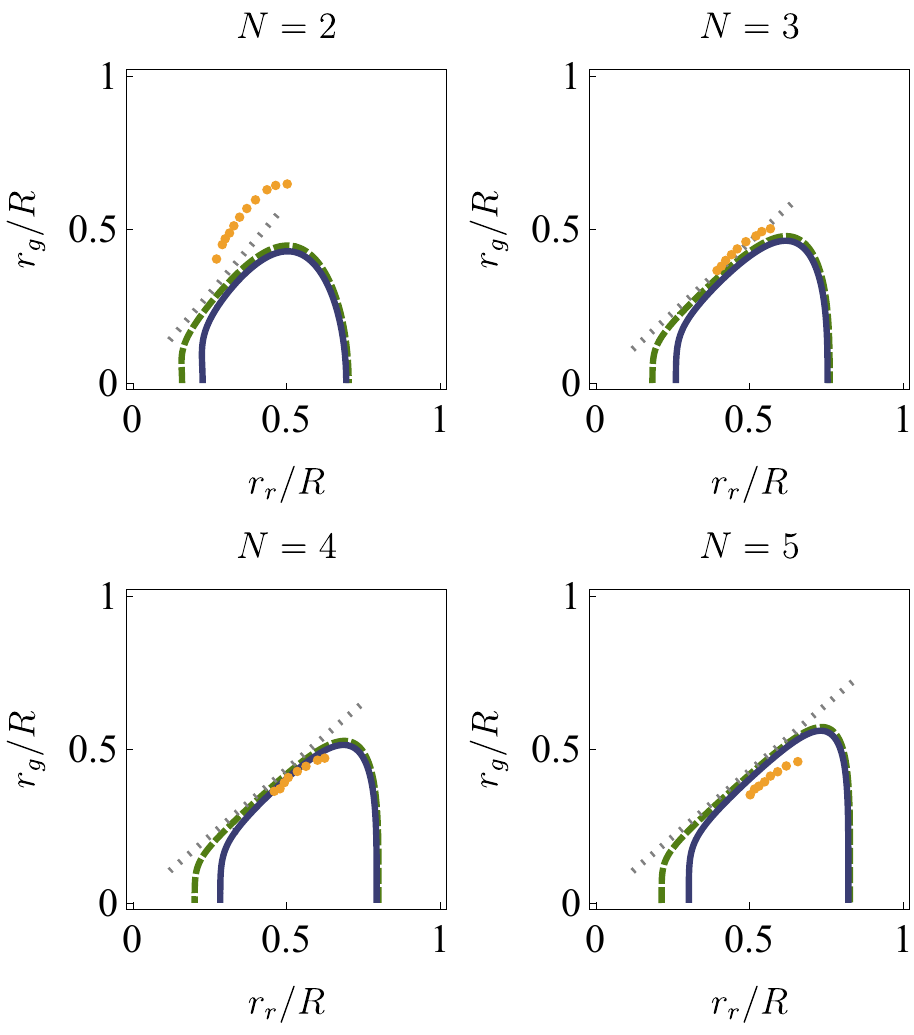}
    \caption{Functional relation between the radius of the $N$-gon of ghost vortices ($r_g$) and that of the $N$-gon of real vortices ($r_r$). The curves were obtained by combining Eqs. (\ref{eq:Precession_frequency_vs_r_r}) and (\ref{eq:r_r_vs_r_g}) for $m_a/m_b=23/39$ (this value is referred to a heteronuclear mixture of ${}^{23}\mathrm{Na}$ and ${}^{39}\mathrm{K}$ \cite{Richaud2019}). Green dashed and blue solid lines correspond, respectively, to $\mu=0.05$, and $\mu=0.10$. Gray dotted lines correspond to the linearized relation (\ref{eq:Linearized_r_r_r_g}), while orange dots correspond to measures extracted from GP simulations The following parameters have been used: $N_a=5 \times 10^4$, $N_b=1.48 \times 10^3$,  $R=50\, \mu m$, $m_a=3.82 \times 10^{-26}$ kg,  $m_b=6.48 \times 10^{-26}$ kg, $g_a=52 \times (4 \pi \hbar^2 a_0)/{m_a}$, $g_b=7.6 \times (4 \pi \hbar^2 a_0)/{m_b}$, $g_{ab}=24.2 \times (2 \pi \hbar^2 a_0)/{m_{ab}}$, $l_z=2\, \mu m$.} 
    \label{fig:r_r_vs_r_g}
\end{figure}

\begin{figure}
    \centering
    \includegraphics[width=1\columnwidth]
    {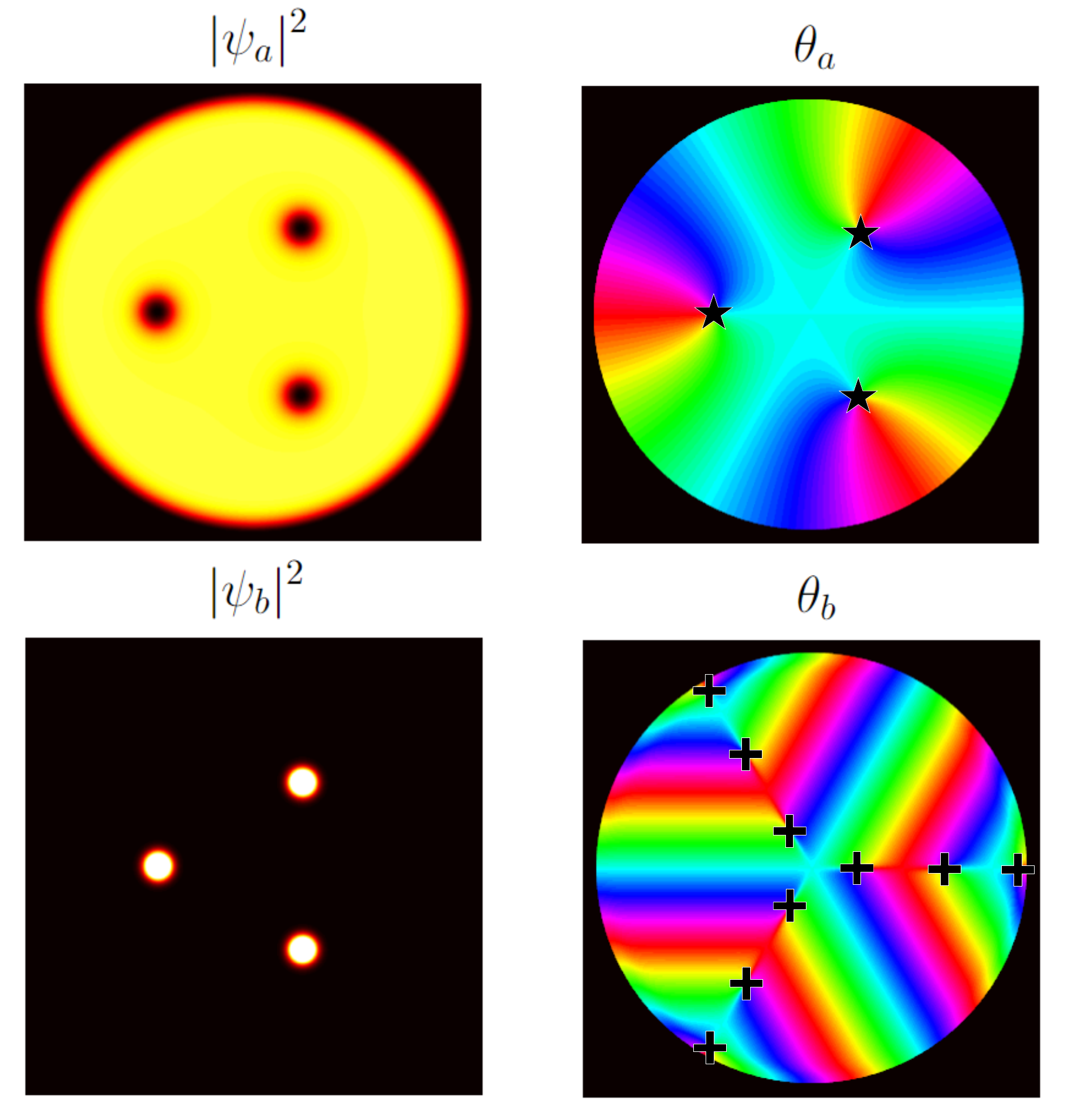}
    \caption{Depending on the ratio $m_a/m_b$ and on the number of real vortices, the slope $r_g/r_r$ of the linearized relation (\ref{eq:Linearized_r_r_r_g}) may turn complex. In this case, the regular $N$-gon-like array of ghost vortices (see, e.g. Fig. \ref{fig:Ghost_vortices_arrangement}) gives way to more complex structures. In the case of a mixture of ${}^{23}\mathrm{Na}$ $+$ ${}^{87}\mathrm{Rb}$ \cite{Wang_2016} atoms and $N=3$, for which relation (\ref{eq:Linearized_r_r_r_g}) would be complex, one can observe the presence of {\it nine} ghost vortices (black crosses) having different radial positions.The following parameters have been used: $N_a=5 \times 10^4$, $N_b=4\times 10^2$, $\Omega=6$ rad/s,  $R=50\, \mu m$, $m_a=3.82 \times 10^{-26}$ kg,  $m_b=1.45 \times 10^{-25}$ kg, $g_a=52 \times (4 \pi \hbar^2 a_0)/{m_a}$, $g_b=100.4 \times (4 \pi \hbar^2 a_0)/{m_b}$, $g_{ab}=73 \times (2 \pi \hbar^2 a_0)/{m_{ab}}$, $l_z=2\, \mu m$.}
    \label{fig:Na_Rb_N_gon}
\end{figure}

Figure \ref{fig:r_r_vs_r_g} illustrates the functional relations between the radius of the $N$-gon of ghost vortices and that of the $N$-gon of real vortices, as predicted by Eqs. (\ref{eq:Precession_frequency_vs_r_r}) and (\ref{eq:r_r_vs_r_g}), together with some results (orange dots) extracted from GP simulations. The resulting function $r_g=r_g(r_r)$ is defined in the interval $[r_{r,1},\,r_{r,2}]$ (with $0<r_{r,1}<r_{r,2}<1$), it is manifestly non-monotonic and has a local maximum at $r_{r,M}$ such that $r_{r,1}<r_{r,M}<r_{r,2}$. By comparing Fig. \ref{fig:Omega_N_vs_r_r} and Fig. \ref{fig:r_r_vs_r_g}, one can notice that the left bound $r_{r,1}$ of the function $r_g(r_r)$ illustrated in Fig. \ref{fig:r_r_vs_r_g} corresponds to $\min_{\Omega_N} r_r$ of Eq. (\ref{eq:Precession_frequency_vs_r_r}), regarded as an implicit function and illustrated in Fig. \ref{fig:Omega_N_vs_r_r}. Interestingly, this circumstance implies that, for a given value of $N$, the smallest possible $N$-gon of real vortices is associated to a \emph{degenerate} $N$-gon of ghost vortices, i.e. to a single ghost vortex of charge $N$ located at the origin. 
Conversely, the right bound $r_{r,2}$ of the function $r_g(r_r)$ illustrated in Fig. \ref{fig:r_r_vs_r_g} does not correspond to $\max_{\Omega_N} r_r$ of Eq. (\ref{eq:Precession_frequency_vs_r_r}). In other words, there exist values of $r_r>r_{r,2}$ for which Eq. (\ref{eq:Precession_frequency_vs_r_r}) admits solutions, but for which Eq. (\ref{eq:r_r_vs_r_g}) does not. In this case, one may wonder how it is possible that the $N$ component-$b$ cores maintain a uniform precession. In fact, if the $N$-gon of ghost vortices no longer exists, what supports the precession of component-$b$ cores? The answer is that, for values of $r_r>r_{r,2}$, the very structure of the phase field $\theta_b$ undergoes profound changes, i.e. it significantly departs from Eq. (\ref{eq:theta_b}), as additional ghost vortices appear and their geometric arrangement is no longer that of a regular $N$-gon (see, e.g., the bottom right panel of Fig.  \ref{fig:Na_Rb_N_gon}). 


The discussed properties of ghost vortices' configurations are robust with respect to small variations of $N_b$ (the number of component-$b$ cores). As visible from Fig. \ref{fig:r_r_vs_r_g}, in fact, the functional relation $r_g(r_r)$ is almost unaffected upon doubling $\mu$ from 0.05 to 0.10. More specifically, an increase of $\mu$ seems only to (slightly) enhance the value of the left bound $r_{r,1}$ of the domain of the function $r_g(r_r)$. This property follows from those of Eq. (\ref{eq:Precession_frequency_vs_r_r}), especially the inequality $\min_{\Omega_N}r_r(\mu_>)>\min_{\Omega_N}r_r(\mu_<)$ for $\mu_>>\mu_<$, as illustrated in Fig. \ref{fig:Omega_N_vs_r_r}. Moreover, all panels of Fig. \ref{fig:r_r_vs_r_g} are marked by the presence of a {\it linear} segment for a rather extended range of values of $r_r$. Neglecting the mass-term $\propto \mu$ and the effect of image vortices in Eq. (\ref{eq:Precession_frequency_vs_r_r}), one can obtain the analytic expression of this linear segment:
\begin{equation}
\label{eq:Linearized_r_r_r_g}
    r_g=r_r\left(\frac{m_a}{m_b}\frac{2N}{N-1}-1\right)^{\frac{1}{N}}.
\end{equation}
One can notice (see gray dotted lines in Fig. \ref{fig:r_r_vs_r_g}) that this linearized functional relation well approximates  the central region of all panels for any value of $\mu$, the slope $r_g/r_r$ being a function only of the {\it atomic-mass ratio} $m_a/m_b$ and of the number of vortices $N_v$. 

As anticipated, the functional relation $r_g(r_r)$ obtained by combining Eqs. (\ref{eq:Precession_frequency_vs_r_r}) and (\ref{eq:r_r_vs_r_g}) was benchmarked against numerical simulations of coupled GP equations (\ref{eq:GPE_a}) and (\ref{eq:GPE_b}). More specifically, we found the minimum-energy state of a heteronuclear mixture of ${}^{23}\mathrm{Na}$ and ${}^{39}\mathrm{K}$ \cite{Richaud2019} in a reference frame rotating at frequency $\Omega$ \cite{Fetter2009}. As customary, we replaced $t\to-i \tau$ in Eqs. (\ref{eq:GPE_a}) and (\ref{eq:GPE_b}) and added an extra term $-\Omega \hat{L}_{z}\psi_a$ ($-\Omega \hat{L}_{z}\psi_b$) to the right-hand side of the former (latter) equation. We recall that $\Omega$ is the angular frequency at which the $N$-gon of real vortices rotates (see Sec. \ref{sub:Rotating_Ngons}) while $\hat{L}_{z}$ is the $z$-component of the angular-momentum operator. Sweeping the parameter $\Omega$, and performing the imaginary-time evolution until convergence, we obtain states featuring $N$-gons of reals vortices in $\psi_a$ and $N$-gons of ghost vortices in $\psi_b$ (see Fig. \ref{fig:Ghost_vortices_arrangement}). We further post-process the output to extract the values of $r_r$ and $r_g$. The result of our extensive numerical investigation corresponds to the set of orange dots in Fig. \ref{fig:r_r_vs_r_g}, which well match the corresponding analytical prevision (lines).

It is also worth mentioning that the linearized relation (\ref{eq:Linearized_r_r_r_g}) offers additional insight into the physics of ghost vortices. For the ${}^{23}\mathrm{Na}$ $+$ ${}^{39}\mathrm{K}$ mixture, in fact, the slope $r_g/r_r$ is always real, as the quantity $(m_a/m_b)/[2N/(N-1)]$ is larger than $1$ for $N_v\ge 2$. This condition ensures the existence of the (linear part of the) relation $r_g(r_r)$ and hence that of the corresponding array of ghost vortices. For mixtures featuring a different ratio $m_a/m_b$, the situation can be dramatically different, as the term $(m_a/m_b)/[2N/(N-1)]$ in Eq. (\ref{eq:Linearized_r_r_r_g}) may be smaller than $1$ for some values of $N$. Consider, for example, the case of the mixture ${}^{23}\mathrm{Na}$ $+$ ${}^{87}\mathrm{Rb}$, for which $m_a/m_b\approx 0.26$. Already for $N\ge 3$, the slope $r_g/r_r$ in Eq. (\ref{eq:Linearized_r_r_r_g}) turns complex, thus signalling the breakdown of the associated $N$-gon-like arrangement of ghost vortices. In these circumstances, in fact, ghost vortices are arranged in more complex and articulated structures. Although a systematic analysis of these structures is beyond the scope of the present work, we illustrate an example where {\it nine} same-signed ghost vortices support the uniform precession of {\it three} real vortices (see Fig. \ref{fig:Na_Rb_N_gon}).

At this point, one may argue that ghost vortices in the wavefunction $\psi_b$ (see the last row of Fig. \ref{fig:Ghost_vortices_arrangement}) could be a sort of ``numerical artefact". After all, the phase field $\theta_b$ illustrated therein is nothing but $\arg(\psi_b)$ and, since $|\psi_b|^2$ is vanishingly small in the neighbourhood of ghost vortices, the very existence of these phase singularities may be questionable. In the next section, we show that this is far from being true and that, instead, these ghost vortices can be made real and hence directly observed by means of a simple quench protocol.

\section{Making ghost vortices real}
\label{sec:GPE_Quenches}
As discussed in Sec. \ref{sec:Ngos_of_ghost_vortices}, due to the irrotational character of the flow (\ref{eq:J_b}) associated to $\psi_b$, the set of massive cores can precess together with their hosting vortices only if the phase field $\theta_b$ includes a suitable set of phase singularities in regions where the density field $|\psi_b|^2$ is vanishingly small. Yet, a viable experimental protocol to \emph{directly} probe their existence is missing and would be desirable.

The quench protocol that we propose is based on the use of an ultracold mixture of ${}^{23}\mathrm{Na}$ and ${}^{39}\mathrm{K}$ atoms (both in the spin state $|f=1,\,m_f=-1\rangle$). This mixture can be doubly Bose-condensed \cite{Schulze2018} and is characterized by the possibility to tune the interspecies scattering length $a_{\mathrm{Na,K}}\propto g_{ab}$ across the miscible-immiscible transition \cite{Richaud2019,Penna2018} via magnetic Feshbach resonances \cite{Chin2010}. In this implementation, the Sodium wavefunction plays the role of $\psi_a$, while that of Potassium corresponds to $\psi_b$. Sodium atoms are assumed to be confined by a box-like circular potential \cite{Navon2021} (which can be realized by means of modern DMD technology), while Potassium atoms are further confined within the component-$a$ vortices when the immiscibility condition $g_{ab}>\sqrt{g_a g_b}$ is satisfied \cite{Richaud2019Supermix}. Both atomic clouds can be made quasi-2D thanks to a strong harmonic confinement along the $z$-axis, its effective thickness corresponding to parameter $d_z$ introduced in Sec. \ref{sec:Rotatig_two_component_BEC} (see also Ref. \cite{Richaud2022Collisions} for more details concerning the experimental preparation of massive-vortices arrays). 

\begin{figure}[h!]
   \centering
   \includegraphics[width=\columnwidth]{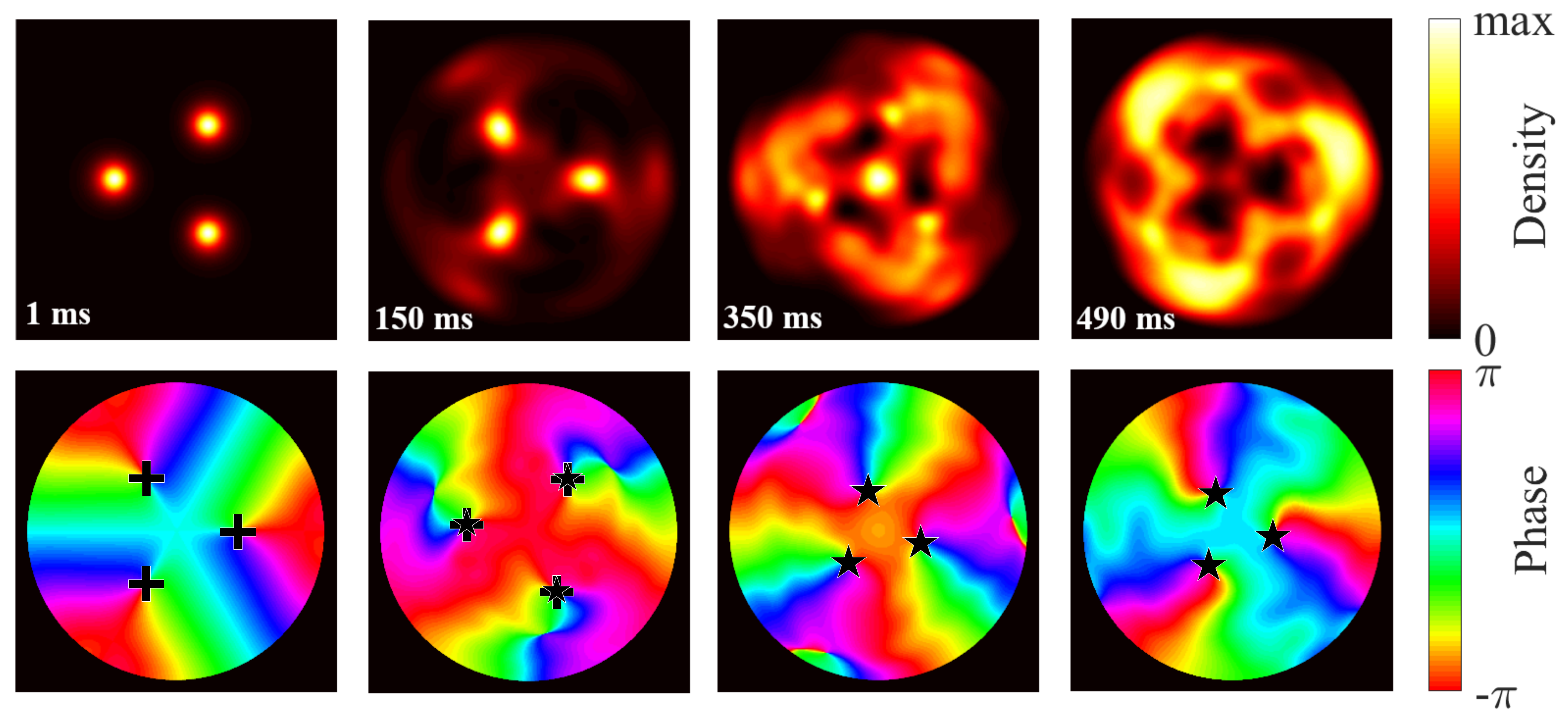}
    \caption{Proposed quench protocol to catch ghost vortices in $\psi_b$. Density $\rho_b$ (first row) and phase $\theta_b$ (second row) associated to the wavefunction of the core-filling component. First column: three component-$b$ cores exhibit precession due to the presence of as many ghost vortices (see black crosses in the phase plots). Second column: upon quenching the interaction parameter $g_{ab}$ across the immiscible-miscible transition, component-$b$ condensate starts to diffuse. Third column: while diffusing, component-$b$ bosons swirl around the ghost-vortices positions. Fourth column: ghost vortices have turned into real vortices (see black stars in the phase plot). The video showing the full dynamical evolution can be found in the Supplemental Material \cite{SM}. The following parameters have been used: $N_a=5 \times 10^4$, $N_b=10^3$, $\Omega=7$ rad/s,  $R=50\, \mu m$, $m_a=3.82 \times 10^{-26}$ kg,  $m_b=6.48 \times 10^{-26}$ kg, $g_a=52 \times (4 \pi \hbar^2 a_0)/{m_a}$, $g_b=7.6 \times (4 \pi \hbar^2 a_0)/{m_b}$, $g_{ab}=24.2 \times (2 \pi \hbar^2 a_0)/{m_{ab}}$, $l_z=2\, \mu m$. }
    \label{fig:quench}
\end{figure}

In analogy with standard fluid-dynamics experiments where colored smoke is used to visualize flow patterns \cite{Sieverding1983}, it is possible to visualize hidden ghost vortices in $\psi_b$ by releasing part of the component-$b$ condensate from the hosting vortices. Figure \ref{fig:quench} illustrates the various stages of the proposed experimental sequence (see also the video in the Supplemental Material \cite{SM}). Starting from an $N$-gon of massive vortices, which are stabilized, for $t<0$, by the assumed immiscibility of the two components, we quench, at time $t=0$, the interaction parameter $g_{ab}\to g_{ab}^\prime=0.5\, g_{ab}$ in such a way that $g_{ab}^\prime<\sqrt{g_a g_b}$, and thus crossing the miscible-immiscible phase boundary  (first column). For $t>0$, part of the component-$b$ bosons diffuses through $\psi_a$ and thus tends to occupy the available space within the circular trap  (second column). Indeed, the linear dispersive (diffusive) term in the GPE for $\psi_b$ supports the spreading of localised wave packets. The crucial observation is that this diffusion process is deeply affected by the presence of ghost vortices, meaning that, component-$b$ bosons, while diffusing out of the original real-vortex cores, give place to a \emph{swirling} flow around the position of each ghost vortex (third column). As a result, the neighborhood of each ghost vortex gets populated by a non vanishing fraction of the component-$b$ BEC and what was a mere phase singularity in an (almost) zero-density region, now turns into a standard vortex (third and fourth column), i.e. a localized density depletion associated to a phase singularity. 

In this perspective, the previously introduced $^{23}\mathrm{Na}$ $+$ $^{39}\mathrm{K}$ mixture constitutes an almost ideal platform, since, varying the applied magnetic field, one can precisely tune the interspecies scattering length $a_{\mathrm{Na-K}}$ while leaving the two intraspecies scattering lengths almost unaffected \cite{Richaud2019}. 

\begin{figure}[h!]
   \centering
   \includegraphics[width=\columnwidth]{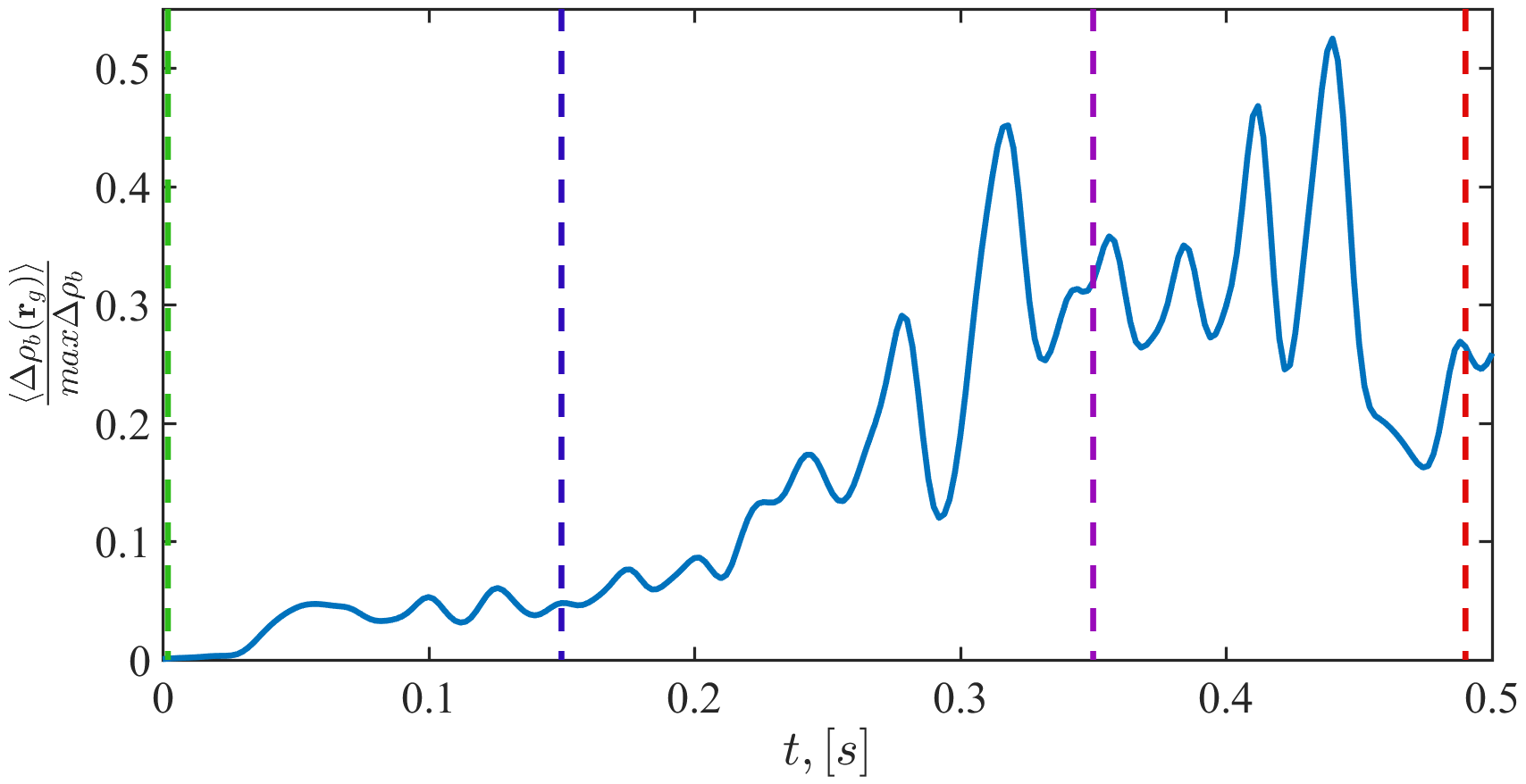}
   \vspace{-0.75cm}
    \caption{A possible indicator to quantify the ``realness" of ghost vortices is the density curvature at vortex position. Here we illustrate the (normalized to the maximum value of the curvature at a given moment) time-dependence of $\Delta\rho_b$ in the neighborhood of ghost vortices. The associated quench protocol is the one illustrated in Fig. \ref{fig:quench} (the assumed microscopic model parameters are listed in the caption thereof). Vertical dashed lines corresponds to the times of the four panels of Fig. \ref{fig:quench}.}
    \label{fig:curvature}
\end{figure}

We conclude by proposing a possible indicator to {\it quantify} the degree of ``realness" (as opposed to ``ghostliness") of a vortex. In analogy to standard computer-vision algorithms used to perform ``edge detection", which are based on Laplacian filters \cite{Haralick1992}, we focus on the \textit{density curvature} $\Delta \rho_b$ in the neighborhood of the phase singularity. Clearly, the density curvature at the center of a purely ghost vortex is zero, while the density curvature at the center of a vortex in a homogeneous BEC reads $\rho_0/\xi^2$ (here $\rho_0$ is the value of the density far from the vortex and $\xi$ the condensate healing length). The result of this analysis is illustrated in Fig. \ref{fig:curvature}. 

One can indeed appreciate the transformation of a {\it ghost} vortex (notice that, at $t=0$, $\Delta\rho_b(\bm{r}_g)$ is vanishingly small) into a {\it real} vortex. As a technical remark, it is worth mentioning that the presented value of $\Delta\rho_b$ has been averaged over a circular neighbourhood centered at $\bm{r}=\bm{r}_g$ of radius $\sim\xi$ and over the $N$ ghost vortices, however after that the plot was quite sharp, so we also smoothed the picture by additional filtering over time. We conclude by mentioning that the scope of the proposed indicator is not limited to GP simulation, as it could be easily applied also for the post-processing of real experimental data, obtained, for example, by absorption imaging.

\section{Conclusions and Outlook}
\label{sec:Conclusions}
In this paper, we demonstrated the existence  of ghost vortices in a superfluid which fills the cores of real vortices of another superfluid and proposed a viable experimental protocol to observe these elusive but intriguing objects.

More specifically, we focused on regular $N$-gons of quantum vortices in $\psi_a$ whose cores  are filled by localized wavepackets of component-$b$ bosons (see Fig. \ref{fig:Ghost_vortices_arrangement}). We started by constructing a fully analytical model which allows to predict the precession frequency of an $N$-gon of equally-signed quantum vortices in the presence of a non-zero core mass [see Eq. (\ref{eq:Precession_frequency_vs_r_r})]. Within this model, we proved that regular $N$-gons of massive vortices are robust configurations, as they constitute energetically-stable states. We also pointed out the existence of a critical value of the cores' mass [see Eq. (\ref{eq:mu_inequality})] above which the $N$-gon configuration breaks down. Then, we shifted the focus from the dynamics of massive vortices to the properties of the core-filling component ($\psi_b$). We showed that the uniform circular precession of the massive cores is necessarily supported by a suitable array of \emph{ghost} vortices in $\psi_b$. Relying on the irrotational character of the velocity field associated with $\psi_b$, we found an analytical expression [Eq. (\ref{eq:r_r_vs_r_g})] connecting the precession frequency of the $N$-gon, the position of real vortices in $\psi_a$ and that of ghost vortices in $\psi_b$. 

We then benchmarked the predictions of the discussed analytical models against numerical simulations of coupled Gross-Pitaevskii equations (see Fig. \ref{fig:r_r_vs_r_g}). While the agreement was remarkable in all those regimes that fall within the validity range of the presented models, our wide-ranging numerical experiments also revealed the existence of more complex structures of ghost vortices (see, e.g. the \emph{double} $N$-gon of ghost vortices illustrated in Fig. \ref{fig:Na_Rb_N_gon}) which typically arise when geometrical and physical constraints forbid the existence of \emph{single} $N$-gons.

We concluded by proposing a detailed experimental protocol to probe the existence of these elusive but topologically required, phase singularities. The such protocol relies on the use of a two-component BEC, namely a $^{23}\mathrm{Na}$ $+$ $^{39}\mathrm{K}$ mixture, which can be conveniently driven across the miscible-immiscible transition \cite{Schulze2018,Richaud2019}. In essence, starting from the immiscible regime, where massive cores are tightly confined within their hosting vortices, one should quench the interaction parameter $g_{ab}$ so as to enter the miscible regime. In this way, part of the component-$b$ fluid is released from the cores and diffuses into $\psi_a$. Crucially, this diffusion comes with the onset of a swirling flow around the ghost-vortices positions and what were mere phase singularities in an (almost) zero-density background get ``dressed" by swirling component-$b$ atoms (see Fig. \ref{fig:quench}). This is how ghost vortices can be turned into real vortices. Eventually, it is worth mentioning that we quantified the degree of ``realness" of a given phase singularity of a wavefunction by proposing a suitable indicator (the curvature of the density field) which is well-known in the context of Computer Vision as a way to detect features \cite{Haralick1992}, and which can be conveniently employed to post-process both numerical-simulations and experimental absorption-images data.   

Our work is expected to open the doors to interesting future developments. As already mentioned, it would be interesting to deepen the study of more complex ghost-vortex configurations, both within a suitable analytical model and by means of numerical experiments. Besides this, another possibility is certainly the investigation of the normal (Tkachenko-like) modes of a massive-vortex $N$-gon (possibly in the presence of mass imbalance \cite{Edmonds2021} or damping \cite{Williamson2021}), which will thus generalize the results of Ref. \cite{Kim2004} obtained in the massless case and further stimulate the current experimental research in real-time vortex dynamics \cite{Serafini2017,Stockdale2020,Kwon2021}. A further possible research direction would point towards the quantum properties of the trapped component, a study which should involve the extraction of an effective (Bose)-Hubbard model where arrays of quantum vortices play the role of effective optical lattices \cite{Chaviguri2017,Chaviguri2018}. This possibility is rather intriguing in relation to the discussed vortex $N$-gons, as the precession motion which they naturally exhibit corresponds to an effective magnetic field for component-$b$ neutral atoms. In this optics, the extremely rich phenomenology associated with Hubbard-like rings pierced by synthetic magnetic fields (see Refs. \cite{Amico2022,Naldesi2022,Chetcuti2022,Richaud2021PRB,Pecci2021,Amico2021,Polo2020,PerezObiol2022}  and references therein) could be recreated (and revisited) in an optical-lattice-free platform.

\section*{Acknowledgements}
The authors are grateful to Luca Salasnich for his discussions and comments about this paper. A.R. received funding from the European Union’s Horizon research and innovation programme under the Marie Skłodowska-Curie grant agreement \textit{Vortexons} no. 101062887, by Grant No. PID2020-113565GB-C21 funded by MCIN/AEI/10.13039/501100011033, and  by grant 2021 SGR 01411 funded by Generalitat de Catalunya. A.Y. acknowledges support from BIRD Project ``Ultracold atoms in curved geometries" of the University of Padova and National Research Foundation of Ukraine through grant No. 2020.02/0032.

\begin{appendices}

\section{Potential energy of a vortex $N$-gon}
\label{app:Potential_energy}
When considering regular $N$-gons of vortices, the effective potential  present in the point-like Lagrangian model can be computed in closed form. The crucial step to perform this computation involves the following summation:
\begin{equation} 
\label{eq:S_1}
  S_1=\sum_{s=0}^{N-1}\ln{\left[1-2x \cos{\left({\frac{2 \pi s }{N}}\right)}+ x^2\right]}. 
\end{equation}
One can compute it by using the generating functions of the Chebyshev polynomials in the following form:
\begin{equation} \label{eq:Chebyshev_generating_function}
    \sum_{n=1}^{\infty} \cos(n\theta) \frac{x^n}{n} =\ln{\left(\frac{1}{\sqrt{ 1-2x \cos\theta+ x^2}}\right)}. 
\end{equation}
Also, one needs the following relation:
\begin{equation} \label{eq:Cos_sum}
    \sum_{s=0}^{N-1} \cos{\left(\frac{2 \pi s}{N} k \right)} = 
    \left\{
    \begin{array}{cc}
           N & \mbox{if} \, \, k = Z N\\
           0 & \mbox{if} \, \, k \neq Z N
    \end{array}
    \right.
\end{equation}
where, $Z$ is an integer number. This expression is valid provided that $k$ is integer and $N$ is natural.

By choosing $\theta=\frac{2 \pi s}{N}$ in the generating function (\ref{eq:Chebyshev_generating_function}), and summing over $s$, we get the following relation:  
\begin{equation} 
    \sum_{n=1}^{\infty} \left[ \sum_{s=0}^{N-1} \cos{\left(\frac{2 \pi s}{N} n \right)} \right]\frac{x^n}{n} =-\frac{1}{2} S_1.
\end{equation}
From Eq. (\ref{eq:Cos_sum}) we see that only $N$-folded terms survive after summation over $s$. The remaining sum over $n$ has the form of the Taylor series expansion for the logarithm. Finally, we obtain:   
 \begin{equation} \label{eq:S_1 final}
            S_1=2\ln(|1-x^{N}|) .
 \end{equation}
Such an approach is correct only for $|x|<1$, due to series convergence. However, one can easily continue for all values of $x$, after a few tricks with variable substitutions for $S_1$.  

Let us now compute the potential energy for the stationary vortex $N$-gon with radius $r$, which has the following dimensionless form:
$$
           V= \sum_{j=1}^N \ln \left(1-r^2\right) 
$$
\begin{equation}
    + \sum_{i<j} \ln  \left[\frac{1 - 2 \cos{ \left(\frac{2 \pi }{N} (i-j) \right)}  r^2 + r^4}{r^2 -  2 \cos{ \left(\frac{2 \pi }{N} (i-j) \right)} r^2  + r^2}\right].
\end{equation}

As the second double summation depends only on the term $\cos{ \left(\frac{2 \pi }{N} (i-j) \right)}$, using its parity and periodicity properties, we can rewrite it into a more convenient form by recalling that:
$$
     \sum_{i<j} f(i-j)=\frac{1}{2} \sum_{j=1}^N \sum_{i=1, i \neq j}^N f(i-j)  
$$
\begin{equation}
    =\frac{1}{2} \sum_{j=1}^N \sum_{s=1}^{N-1} f(s)=\frac{N}{2} \sum_{s=1}^{N-1} f(s)  
\end{equation}
Thus $V$ can be recast as:
$$
             V=N\ln \left(1-r^2\right) + \frac{N}{2}\sum_{s=1}^{N-1} \ln {\left[1-2 \cos{ \left(\frac{2\pi}{N} s \right)} r^2 +r^4 \right]}  
$$
\begin{equation}
    -N\sum_{s=1}^{N-1} \ln{(r)} -\frac{N}{2}\sum_{s=1}^{N-1} \ln {\left[1-2 \cos{ \left(\frac{2\pi}{N} s \right)} + 1\right]}
\end{equation}
The first summation corresponds to $S_1$ [see Eq. \eqref{eq:S_1}], but without the first term at $x=r^2$. The last one corresponds also to $S_1$ \eqref{eq:S_1} without the first term, but here at $x=1$, so one can get the closed form, by taking the limit in the following way:
$$
      \sum_{s=1}^{N-1} \ln {\left[1-2 \cos{ \left(\frac{2\pi}{N} s \right)} + 1\right]}  
$$
\begin{equation}
    = \lim_{x \xrightarrow{} 1^{-}} 2\left[\ln{(1-x^N)}- \ln{(1-x)}\right]=2\ln{(N)}
\end{equation}
Finally, upon substituting we get the potential energy $V$:
\begin{equation}
        V=N\left[ \ln(1-r^{2N}) - (N-1) \ln{(r)}-\ln{(N)}\right]
\end{equation}
Thus one can get the following Hamiltonian in the rotating frame for the $N$-gon:
$$
      H'_N= N \left[\ln{(1-{r}^{2N})}-(N-1)\ln{(r)}  \right. 
$$
\begin{equation}
    \left. - \ln{(N)}
       -\ln{(\xi_a)} -\Omega (1-{r}^2) - \frac{\mu}{2N}{r}^2 \Omega^2\right] 
\end{equation}
If an extra central vortex is present, the total potential energy of the necklace reads:
\begin{equation}
    H_{N+1}=H_N + H_{0}
\end{equation}
where
\begin{equation}
    H_N=N\left[\frac{\mu}{2(N+1)}\Omega^2 r^2 - \ln{(\xi_a)} + \ln\left(\frac{1-r^{2N}}{N\,r^{N-1}} \right)\right]
\end{equation}
corresponds to the $N$ vortices constituting the $N$-gon and 
\begin{equation}
    H_0 = -2 N\ln {(r)} - \ln{(\xi_a)}
\end{equation}
is due to the central vortex. The angular momentum associated to this configuration reads $\bm{l}=\bm{l}_0+\sum_{j=1}^N \bm{l}_j$, where 
\begin{equation}
    \bm{l}_j\cdot\hat{z}= 1-r^2 + \frac{\mu}{N+1}\Omega r^2
\end{equation}
for $j=1,\,\dots,\,N$ and
\begin{equation}
    \bm{l}_{0}\cdot\hat{z}= 1
\end{equation}
for the central vortex.
So, the total energy in the rotating reference frame reads 
\begin{equation}
    H_{N+1}^\prime = H_{N+1}-\Omega \left( \bm{l}_0 +\sum_{j=1}^N \bm{l}_j \right)\cdot\hat{z}.
\end{equation}

\end{appendices}

\end{document}